\documentclass[12pt]{article}
\UseRawInputEncoding 
\usepackage[hmargin=2.5cm,vmargin=3cm]{geometry}
 \usepackage{graphicx}
 \usepackage{amsmath}
 \usepackage{amssymb}
  \usepackage{enumerate}
\usepackage{cite}
\usepackage{mathabx}
\newcommand{\bey}{\begin{eqnarray}}
\newcommand{\eey}{\end{eqnarray}}

\begin{document}
\title{ Quantum Information in Relativity: the Challenge of QFT Measurements } \author {Charis Anastopoulos\footnote{anastop@physics.upatras.gr}  \;   and   \; Ntina Savvidou\footnote{ksavvidou@upatras.gr}\\
 {\small Department of Physics, University of Patras, 26500 Greece} }

\maketitle

\begin{abstract}
Proposed quantum experiments in deep space will be able to explore quantum information issues in regimes where relativistic effects are important.
In this essay, we argue that a proper extension of Quantum Information theory into the relativistic domain requires the expression of all informational notions in terms of quantum field theoretic (QFT) concepts. This task requires a working and practicable theory of QFT measurements. We present the foundational problems in constructing such a theory, especially in relation to longstanding causality and locality issues in the foundations of QFT. Finally, we present the ongoing Quantum Temporal Probabilities program for constructing a measurement theory that (i) works, in principle, for any QFT, (ii) allows for a first-
principles investigation of all relevant issues of causality and locality,
and (iii) it can be directly applied to experiments of current interest.

\end{abstract}

\section{Introduction}

Quantum theory and General Relativity (GR) are the two main pillars of modern theoretical physics. Each theory is highly successful in its domain. However, they are structurally incompatible. For example, in quantum theory, the concept of measurement appears to be fundamental part of the formalism, while measurements in GR are derivative concepts. Time in GR is dynamical, while in quantum theory it is described as an external  parameter. Finding a unifying theory, a theory of quantum gravity, is one of the most important goals of current research.
However, there is no functional theory of quantum gravity. This is largely due to the lack of experimental data to guide the theory. Quantum gravitational phenomena are estimated to be significant at the Planck scale, which is well outside our  experimental reach, at least directly.

A large part of quantum gravity research focusses on quantum phenomena at black holes, mainly because the emergence of black hole thermodynamics is expected to be an important feature of any quantum gravity theory. In recent years, strong emphasis has been given on the properties of entanglement in this context (e.g.,  \cite{AMPS}), in order to relate black hole entropy to entanglement entropy \cite{Solo} and to address the problem of information loss in black holes   \cite{InfLo}.

This development seemingly signals a convergence of quantum gravity research and quantum information theory (QIT). After all, entanglement is
the primary motive force behind explosive developments of QIT in the last quarter of a century; it is  the crucial  resource for quantum computing, quantum metrology, quantum communication and more.

This convergence is actually illusory. Entanglement may be a well defined mathematical quantity in Quantum Field Theory (QFT) \cite{entent, HoSa18}, but its status as an informational quantity / resource must be embedded within a comprehensive Quantum Information theory (QIT). However, this is not the case. So far, QIT has been largely developed in the context of non-relativistic quantum mechanics, a small corner of full QFT.  It is ostensibly inadequate when basic relativistic principles – both special and general -- such as causality and covariance, need be accounted for.


Furthermore, entanglement and other resources of QIT \cite{ChGo} refer to the properties of the quantum state at one moment of time. By definition, they cannot account for information that is contained in multi-time correlations. This is a severe limitation in its application to black holes, because   any discussion of the informational balance  in the process of black hole formation and evaporation must take the existence of multi-time correlations into account \cite{QTP6}.

In this paper, we contend that  the proper union of QIT and QFT requires a {\em first-principles} analysis. Information must be defined in terms of the ways that it can be extracted from a quantum system, and this necessitates an analysis of measurements in QFT that goes beyond the current  state of the art. Furthermore, a spacetime-covariant QIT must treat temporal correlations on the same footing with spatial correlations, and it must fully incorporate the principles of causality and covariance. This is the motivation for the Quantum Temporal Probabilities (QTP) program, in development by our team \cite{QTP1, QTP2, QTP3},  that aims to construct a general quantum measurement theory based solely on QFT, and then, to use this as a basis of a relativistic QIT.

Building a sound theoretical foundation for relativistic QIT is not just for the sake of theoretical completeness. Relativistic effects are indispensable in hyper-sensitive  quantum experiments in space to account for the effects of motion (relative velocity, acceleration, or rotation) and gravity on quantum resources \cite{DSQL, Rideout}.

Furthermore, recent research led to the realization that there is a new class of phenomena that can provide novel direct information about the coexistence of gravity and quantum theory. The usual estimate that quantum gravity effects become important at the Planck length-scale, $L_p = 1.6 \times 10^{-35} m$, follows from the assumption that particles manifest quantum behavior at length scales of the order of their de Broglie wavelength. This assumption holds only for a subset of quantum states, relevant to a specific class of experiments, for example, particle scattering experiments.

It is now possible to prepare particles in states that manifest quantum behavior at mesoscopic or even macroscopic scales---see, for example, \cite{ArHo14}. Schr\"odinger cat states are an example of such states, i.e., quantum superpositions of localised states  for particles of mass $M$, with a macroscopic distance $L$ of  their centers. For such states, the effect of gravity becomes stronger as $L$ and $M$ increase \cite{AnHu15}. The search of gravitational effects in such states is now possible, and this raises novel theoretical issues about the interplay of gravity and quantum, especially in relation to locality, causality and information \cite{MQPG}. In this field also, the development of a relativistic QIT is crucial.

  We contend  that a relativistic QIT must be based on the information content of the probability distributions pertaining on measurement on quantum fields. To this end we need a general and practicable theory of QFT measurements. The prototype of such a theory is Glauber’s photodetection theory \cite{Glauber} that has been immensely successful in quantum optics.  However, Glauber’s theory has a restricted domain of applicability (photons), and  it faces problems with causality in set-ups that involve photons traveling long distances before measurements.

The QTP method---which we intend as a primary vehicle for incorporating QIT notions in QFT---is an improvement over Glauber's theory.
 The idea of QTP is  to move beyond the description of quantum theory in terms of single-time quantum states and rely on the notion of histories. The simplest example of a history is a sequence of properties (measurement outcomes) of a physical system at different moments of time, but they can also describe time-extended properties of the system, which are essential for the formulation of a relativistic quantum information theory.

A well-known histories formulation of quantum theory is the decoherent histories framework that has been developed by Griffiths, Omnés, Gell-Mann and Hartle \cite{histories}. They represent histories by strings of projection operators at different moments of time, and they define probabilities for sets of histories that satisfy a decoherence condition. The Histories Projection Operator   theory  provides a mathematically rigorous generalization for decoherent histories that allows for the description of continuous time \cite{HPO}.  A  histories theory, developed by one of us (K.S.)\cite{Savvidou, Sav09}, incorporated a novel temporal structure into HPO theory, and made it possible to define spacetime-extended quantum observables.

QTP employs the conceptual and mathematical tools of histories theory in order to express quantum probabilities in terms of QFT correlation functions. The probability density associated with n measurement events is a linear functional of a 2n unequal-time field correlation function. The QTP method leads to probabilities in which the spacetime point is treated as a random variable, i.e., the observables are time extended. This property is absent from past quantum measurement formalisms that were designed for non-relativistic quantum theory. It provides a more accurate representation of particle detection, and it connects straightforwardly with the familiar formulations of QFT through functional methods.

\smallskip

The structure of this paper is the following. In Sec. 2, we point out the main differences between QFT and QIT in relation to causality and locality. In Sec. 3, we analyse the difficulty in describing local measurements in QFT, which is the origin of all difficulties in defining QIT concepts. In Sec. 4, we present some ideas and models on QFT measurements, and in Sec. 5 we present  the QTP program, and its status.

\section{Current incompatibilities between QIT and QFT}
In this section, we briefly describe the structures of QFT and QIT, in order to highlight their current incompatibility.

 QFT is a quantum theory that incorporates additional principles about the effect of spacetime structure on the properties of quantum systems \cite{Weinberg, Haag, Duncan}, with  emphasis on  the causal propagation of signals. Among the principles of QFT, the following are particularly relevant to the present discussion.
 \begin{enumerate}
 \item Observables are expressed in terms of quantum field operators $\hat{\Phi}_a(X)$, where $X$ is a point of Minkowski spacetime and $a$ a label that includes both spacetime and internal indices.

 \item The quantum fields transform covariantly under a unitary representation of the Poincar\'e group.  The generators of the Poincar\'e group are local functionals of the fields.

 \item The Hamiltonian, i.e., the generator of time translations,  has strictly positive spectrum. There is a unique ground state, the vacuum, that is invariant under Poincar\'e transformations.

     \item The spacetime causal structure is incorporated into the physical description through the {\em microcausality axiom}. For $X$ and $X'$ spacelike separated points, either $[\hat{\Phi}_a(X), \hat{\Phi}_a(X')] = 0$ or $[\hat{\Phi}_a(X), \hat{\Phi}_a(X')]_+ = 0$. The first case (commutator) corresponds to bosons and the second (anticommutator) corresponds to fermions. Note that the microcausality axiom
          is not necessarily equivalent  to the statement that observables in spacelike separated regions commute.
 \end{enumerate}

QFT is usually formulated in the framework of perturbative S-matrix theory, where the main elements are time-ordered correlation functions. This formulation is useful for most applications, however, it is not mathematically rigorous, and this leads to difficulties in proving important results (e.g., the spin-statistics theorem or the CTP theorem) in full generality. For this reason, the S-matrix formulation is complemented by axiomatic frameworks that purport to derive QFT properties from a set of axioms. The principles given above best fit Wightman's axiomatization \cite{Wightman}, where quantum fields are expressed as operator-valued distributions on a Hilbert space. However, they can easily be implemented in the more general axiomatization by Haag and Kastler \cite{Haag} that formulates QFT through $C^*$-algebras, and also in the language of S-matrix theory\footnote{In the S-matrix description of QFT, locality is incorporated primarily in the {\em cluster decomposition}property of the S-matrix. Cluster decomposition is a property of a specific hierarchy of correlation functions $G_n(X_1, X_2, \ldots, X_n)$ defined by the $S$ matrix; $n = 0, 1, 2, \ldots$. It asserts that
\bey
G_{n+m}(X_1, \ldots, X_n, X'_1, \ldots, X'_m) =  G_n(X_1, X_2, \ldots, X_n) G_n(X'_1, X'_2, \ldots, X'_m)
\eey
 if the cluster of points $X_1, \ldots, X_n$ is spacelike separated from the cluster $X'_1, \ldots, X'_m$.
 Cluster decomposition follows from the locality of the Hamiltonian and the microcausality assumption. Microcausality is also needed for the unitarity of the S-matrix. See, sections 4.4 and 3.5 in Ref. \cite{Weinberg}. }.

If we compare the principles of QFT, in any of its formulations,  with the usual axioms of quantum theory (for example, \cite{despagnat, Isham}), we will notice a glaring absence. No QFT framework  contains a rule of state update after measurement, i.e., a rule for `quantum state reduction'! This is a highly unsatisfactory state of affairs because no probabilistic theory is complete without such a rule.

There are two reasons for the omission of a state update rule in QFT. First,  the usual rule of state reduction for non-relativistic physics is problematic in relativistic setups---see the next section---, and no replacing rule has yet been developed that works in full generality.
Second, most  QFT predictions involve set-ups with a single state preparation and a single-detection event, and they can be described in terms of the S-matrix with no need for a state update rule.
In particular,  cross-sections in high-energy scattering experiments are obtained from S-matrix amplitudes; the  spectrum of composite particles, e.g., hadrons, is determined by S-matrix poles; decay rates of unstable particles are determined from the imaginary part of S-matrix poles\footnote{However, if the decay rates are not constant, i.e., in non-exponential decays, a proper measurement theory is needed in order to construct a positive probability density for the  decay time \cite{Ana19}.}.

On the other hand, in quantum optics, we need joint probabilities of detection in order to describe phenomena that involve higher order coherences of the EM field, like photon bunching and anti-bunching \cite{QuOp}. A first-principles calculation of joint probabilities for multiple measurements is impossible without a state-update rule. In practice, joint probabilities are expressed in terms of photodetection models, like Glauber's, whose derivation is rather heuristic and it avoids explicit state updating. However, planned experiments in deep space \cite{Rideout, DSQL} that involve  measurement of EM field correlations will arguably require a first-principles analysis of joint probabilities in order to take into account  the relative motion of detectors and delayed propagation at long distances.

QIT is not a closed theory that can be brought into an axiomatic form, rather it is a set of ideas, techniques and method that explores the informational properties of quantum systems. There is no intrinsic limitation why QIT cannot be applied to relativistic systems, but historically its methods originate from non-relativistic quantum theory. The most important set of methods is provided by the  Local Operations and Classical Communication (LOCC) paradigm \cite{LOCC}, which provides a concrete implementation of the notions of locality and causality in QIT. The Hilbert space of any informational system is split us a tensor product
$\otimes_i {\cal H}_i$, where ${\cal H}_i$ is the Hilbert space of the $i$-th subsystem. A local operation on the $i$-th subsystem is a set of completely positive maps ${\cal C}^{(i)}(a)$  on states of ${\cal H}_i$, such that $\sum_a {\cal C}^{(i)}(a) = \hat{I}$; here, $a$ are the measurement outcomes of the operation. In some abstract frameworks, local operations   constitute the events in quantum systems \cite{causalQI}.

Causality is implemented through the concept of classical communication. An operation ${\cal C}^{(i)}(a)$ on a subsystem $i$ may depend on the outcome $b$ of an operation ${\cal D}^{(j)}(b)$, if the outcome can be communicated to $i$ through a classical channel, prior to the operation ${\cal C}^{(i)}(a)$.  As such, QIT carries the causal structure of classical communication, which is taken as external to the system. QIT is usually not concerned with    real-time quantum signal propagation between disconnected subsystems, which, after all, can be implemented consistently only in a relativistic theory. Causal correlations exist along timelike directions, while quantum correlations studied in QIT are spacelike, in the sense that they are  defined with respect to the single-time properties of the quantum state. Certainly, there exist non-classical temporal correlations in quantum systems\footnote{They are identified by the violation of the Leggett-Garg inequalities \cite{LeGa}, or the violation of Kolmogorov additivity for probabilities of multi-time measurements \cite{Ana06}, a property sometimes referred to as non-signaling in time \cite{Brunnt}.}. However, a unified theory of all correlations in relation to their spacetime character (timelike vs spacelike) is currently missing. We believe that such a unified perspective requires a QFT treatment of quantum information, starting with an analysis of measurements.

 The causal structure of a quantum informational system is specified by the lattice ${\cal L}$ of all events. This consists of elements  $X_1, X_2, \ldots, X_n$, where events (local operations) take place. ${\cal L}$ is equipped with the operation
 $\leq$: we say that $X_1 \leq X_2$ if a classical signal from $X_1$ can reach $X_2$.

 For a relativistic system, one could   identify the elements of ${\cal L}$ with points of Minkowski spacetime $M$ and     $\leq$ with the usual spacetime causal ordering: $X_1 \leq X_2$ if $X_2$ in the future light-cone of $X_1$. Nonetheless, we have to keep in mind spacetime points in QFT serve an additional function: they appear   as arguments of the quantum fields and they are essential to the implementation of dynamics through the representation of the Poincar\'e group\footnote{The Poincar\'e group in QFT plays a double role as a symmetry of both dynamics and of the causal structure. In standard QFT, this distinction is conceptual. However, when writing QFT as a histories theory, it is possible to define two mathematically distinct representations of the Poincar\'e group, one associated with the causal structure and one associated to the dynamics \cite{Savvidou, Sav02, Sav09}.}. A relativistic QIT that respects causality ought to express all operations / measurements on the quantum system in terms of quantum fields. The lack of such a representation is the main limitation of existing approaches to relativistic QITs \cite{Peres2, PoVai, InstMeas, GoPr}.


\section{Problems in describing measurements in QFT}

In the previous section, we saw that the main challenge in the development of a relativistic QIT is the description of measurements / operations in a way that is compatible with locality and causality  In this section, we explain why the description of measurements in non-relativistic quantum theory cannot be transferred to relativistic QFT.

\subsection{Non-covariance of projection rule}
	 It has long been known that the quantum state is genuinely different when recorded from different Lorentz frames in set-ups that involve more than two quantum measurements \cite{RelRed}. To see this, consider a description of a quantum system in Minkowski spacetime, with respect to
an  inertial reference frame
 $\Sigma$. Let the system be prepared in a state   $|\psi\rangle$. The event A corresponds to the measurement of  observable $\hat{A} = \sum_n a_n \hat{P}_n$, where $a_n$ are the eigenvalues of $\hat{A}$ and $\hat{P}_n$ the associated projectors. The event
  B corresponds to the measurement of observable  $\hat{B} = \sum_m b_m \hat{Q}_m$, where $b_m$ are the eigenvalues of $\hat{B}$ and $\hat{Q}_m$ the associated projectors. We assume that the two events are spacelike separated.

  Let the outcome of the two measurements be $a_n$ and  $b_m$, respectively. According to the usual state-update rule, the quantum state evolves as follows.
 \begin{eqnarray}
 \begin{array}{cc} |\psi\rangle& t < t(A)\\
c_1 \hat{P}_n|\psi\rangle & t(A) <t < t(B) \\
 c_2 \hat{Q}_m\hat{P}_n|\psi\rangle& t > t(B), \end{array}
 \end{eqnarray}
where $c_1, c_2$ are constants. This evolution of the state is depicted in Fig. 1.a

Let us now describe the same process in a different reference frame $\Sigma'$. The initial state, the observables and the associated spectral projectors must be transformed to this frame; we denote the transformed quantities by a prime. This transformation is implemented by a unitary representation of the Poincar\'e group, but this fact is irrelevant to present purposes.
In $\Sigma'$,   the quantum state evolves as follows.
 \begin{eqnarray}
 \begin{array}{cc} |\psi'\rangle& t' < t'(A)\\
c_1' \hat{Q}'_n|\psi'\rangle & t'(A) < t' < t'(B) \\
 c_2' \hat{Q}'_m\hat{P}'_n|\psi'\rangle& t' > t'(B), \end{array}
 \end{eqnarray}
where $c_1', c_2'$ are constants. This evolution is depicted in Fig. 1.b.

\begin{figure}[tbp]
\includegraphics[height=4cm]{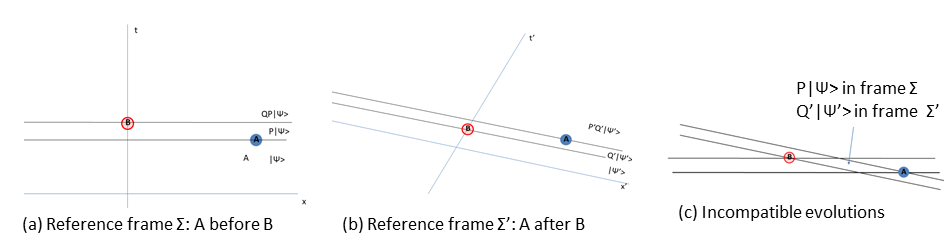} \caption{ \small Evolution of the quantum state under successive measurements in a relativistic system.    (a) Reference frame $\Sigma$. (b) Reference frame $\Sigma'$. (c) The spacetime region in which the two evolutions are incompatible.
}
\end{figure}

The two evolutions give incompatible results in the parallelogram that is indicated in
Fig. 1.c. In this spacetime region, the quantum state is
$c_1 \hat{P}_n|\psi\rangle$ in the reference frame $\Sigma$, and   $c_1' \hat{Q}'_n|\psi'\rangle$  in the reference system $\Sigma'$.  Since   $\hat{X}$ and $\hat{Y}$ are arbitrary, there is no transformation that depends only on the reference frames that take one state to the other. We obtain {\em genuinely different evolutions for the quantum state in the two reference frames}.

Nonetheless, this ambiguity in the quantum state does not lead to an ambiguity in physical predictions, which are expressed in terms for probabilities. The joint probability for the two measurement events is uniquely defined as $\langle \psi|\hat{P}_m\hat{Q}_n|\psi\rangle$, provided that the two observables commute, $[\hat{A}, \hat{B}] = 0$.

The ambiguity in the evolution of the quantum state has led Wigner and others \cite{WH65} to consider the probability rule for multi-time measurements and not the quantum states as fundamental notions. This line of thought  eventually led to formulations of quantum theory in which the fundamental objects are histories of a quantum system \cite{histories} rather than single-time states that evolve in time. Then, probabilities are encoded in the so-called {\em decoherence functional}, a bilinear function in the space of histories.

From the QIT perspective the loss of the quantum state is problematic. Crucial notions like entropy and entanglement are defined in terms of the quantum state, hence, they also share this ambiguity. Furthermore, if the notion of the quantum state is not fundamental, then the mathematical objects that represent the external interventions on the system  should not be defined as completely positive maps acting on single-time states. Rather, they should be defined at the level of multi-time measurements / histories and their probability rules.

\subsection{Spatial localization apparently conflicts causality}
In QIT, the notion of a local quantum system is essential. The natural interpretation of localization in QFT is to consider a quantum system that is localized in a spatial region at a moment of time. However, this notion leads to conflicts with causality, as it is shown by a number of theorems.

Malement's theorem \cite{Malament} asserts that it is impossible to define localization observables, i.e., projectors for a quantum system $\hat{P}_{\Delta}$ that correspond to a spatial region $\Delta$, in a way that
is compatible with Poincar\'e symmetry and causality. Note that Malament's theorem is usually interpreted in terms of particle position observables, but it actually holds for any localized observable.

Theorems from Schlieder \cite{Schlieder} and Hegerfeld \cite{Heg1} show that existing definitions of localizing observables  conflict   the requirement of relativistic causality. Assume, for example, that localization is defined with respect to some spatial observable ${\pmb x}$, leading to  a sufficiently localized probability distribution $\rho({\pmb x}, t)$ for ${\pmb x}$ at some moment of time $t$. Then, time evolution  leads to a probability distribution $\rho({\pmb x}, t')$ that evolves superluminally at latter times $t'$.

The most well known set-up where localization appears to contradict causality is Fermi's two-atom problem. Fermi studied the propagation of information through quantum fields in a system of two localized atoms  \cite{Fermi}. He  assumed that at time $t = 0$,  atom A is in an excited state and atom B in the ground state. He asked when B will notice A and leave its ground state. In accordance with Einstein locality, he found that this happens only at time greater than $r$. It took about thirty years for Shirokov to point out that Fermi's result is an artifact of an approximation \cite{Shiro}.

Several studies followed with conclusions depending on the approximations used. It was believed that non-causality is due to the use of bare initial states, and that it would not be present in a renormalised theory. However, Hegerfeldt showed that non-causality is generic \cite{Heg,Heg2}, as it depends only on the assumption of energy positivity and on the treatment of atoms as localized in disjoint spatial regions------see, also the critique in \cite{Buch} and a recent exactly solvable model \cite{RelQOp2}.

 The localization problem is not an artefact of a particle description, it holds  irrespective of whether one employs particle or field degrees of freedom.
   It is a fundamental issue of QFT that pertains to the definability of local observables and the meaning of locality in relation to quantum measurements.
 It has been
 recognized that localization observables should not be viewed as attributes of particles (or even of their associated  fields),
  but as attributes of the interaction between particles (or fields) and a measuring apparatus \cite{Haag, PeTe}. In this perspective, a solution to the localization problem requires
 a consistent  quantum measurement theory for   relativistic QFTs.

\subsection{Ideal measurements lead to violation of causality}
Sorkin has presented a scenario, in which the existence of ideal measurements in QFT leads to a conflict with locality \cite{Sorkin}. The state-update associated to an ideal measurement transmits information faster than light. The idea is to consider operations on three spacetime regions, $O_1$, $O_2$ and $O_3$, such that $O_2$ is a Cauchy surface, $O_1$ is in the past of $O_2$, $O_3$ is in the future of $O_2$, but $O_1$ and $O_3$ are spacelike separated. In particularly, one acts with a unitary operator on $O_1$, one measures a global observable (corresponding, for example, to subspaces of the QFT Hamiltonian), and then one local observable at $O_3$. The probabilities for the measurement outcomes at $O_3$ depend on the intervention at $O_1$, even if $O_1$ and $O_3$ are spacelike separated.

To avoid this problem  one has to abandon the notion of ideal measurement in QFT and its associated rule of state update. The problem here is that the notion of ideal measurement is essential in the formulation of QIΤ. The maximum amount of information that can be extracted from quantum systems correspond to ideal measurements. In fact, the very notion of a qubit depends implicitly on the accessibility of ideal measurements.


\section{QFT measurement models}
We argued that the development of consistent relativistic QIT requires a measurement theory that respects causality and locality while being expressed in terms of quantum fields. Furthermore, this measurement theory ought to be practical, i.e., it should  lead to non-trivial predictions for setups that are accessible now or in the near future.
In this section, we will give a brief overview of existing models of QFT measurements.

The earliest discussion of QFT measurements was by Landau and Peierls \cite{LP31}, who derived an inequality for the localization of particles. Bohr and Rosenfeld criticized some of their assumptions \cite{BoRo}, and proved the crucial result that the measurement of field properties  requires a test particle of macroscopic scale, in the sense that its charge $Q$ must be much larger than the charge quantum $e$. This analysis implies a distinction with no classical analogue: we will call a microscopic physical system  that interacts with a quantum field {\em a probe} but it is too small to directly measure field properties and leave a macroscopic record (for example, an electron or an ion)\footnote{A probe can be measured by a macroscopic apparatus after it has interacted with the field, and thus provide information about the field.}. We will call {\em detector} any system that can record field properties. By the Bohr-Rosenfeld analysis and by a later theorem of Yanase \cite{Yanase}, any detector must be  a macroscopic system.

The first explicit model for QFT measurement was Glauber's photodetection theory \cite{Glauber}. The theory was developed as a quantum generalization of the classical theory of coherence for the EM field. It  expresses unnormalized probabilities\footnote{The probabilities are unnormalized because most photons in the initial state escape detection.} for photon detection in terms of the electric field operators ${\bf E}(X) $ and the field state $|\psi\rangle$. The (unnormalized) probability density $P(X)$  that a photodetector of photons with polarization parallel to the vector ${\bf n}$ records a photon at spacetime point $X$ is given by
\bey
P(X) =  \langle \psi| E^{(-)}(X) E^{(+)}(X)|\psi\rangle,
\eey
where $E^{(+)}$ is the positive frequency component  and $E^{(-)}$ the negative frequency component of the projected field vector field ${\bf n}\cdot{\bf E}(X)$. Similarly, the joint probability density $P(X_1, X_2)$ for one photon detected at $X_1$ and another at $X_2$ is given by
\bey
P(X_1, X_2) =  \langle \psi| E^{(-)}(X_1)E^{(-)}(X_2)  E^{(+)}(X_2) E^{(+)}(X_1)|\psi\rangle.
\eey
The joint detection probability of photons at different moments of time is essential for the definition of higher order coherences of the electromagnetic field, and for describing  phenomena like the Hanbury-Brown-Twiss effect,   photon bunching and anti-bunching \cite{QuOp}.
The expressions above were originally suggested by the form of the leading-order terms in perturbation theory for the interaction of the EM field with matter. They were not meant to be universal, merely to model the behavior of a general class of photodetectors.

Glauber's theory has been  immensely successful in quantum optics. While it originally refers to photons, its analogues can be constructed for all types of relativistic fields. Its main limitation is that the field splitting into positive and negative frequency components is non-local, and it follows from the so-called Rotating Wave Approximation (RWA) for the interaction of the field to the detector. The RWA
misrepresents the retarded propagation of the electromagnetic field, and for this reason Glauber's theory may face problems with causality in set-ups that involve photons traveling long distances before measurements.

A very common class of models employed for QFT measurements are the {\em Unruh-deWitt} (UdW) detectors \cite{Unruh76, Dewitt}. They first appeared in the study of the Unruh effect, in order to clarify the physical properties of the field that are  experienced by accelerated observers. In an Unruh-deWitt detector, the quantum field is coupled to a point-like system that moves along a pre-determined spacetime trajectory $X(\tau)$, where $\tau$ is the trajectory proper-time. The Hamiltonian is of the form $\hat{H}_{0}\otimes \hat{I} + \hat{I} \otimes \hat{h} + \hat{H}_I$, where $\hat{H}_{0}$ is the field Hamiltonian and $\hat{h}$ is the Hamiltonian of the detector. The most general form of the interaction Hamiltonian is
\bey
\hat{H}_{I} = \hat{O}(X(\tau)) \otimes \hat{m},
\eey
where $\hat{O}(X(\tau))$ is a scalar composite operator for the field, and $\hat{m}$ a self-adjoint operator on the detector Hilbert space.

The  UdW coupling has been used to model the interaction of both probes and detectors with quantum fields, sometimes with misleading terminology. The crucial difference is, as it had been pointed out by Bohr and Rosenfeld \cite{BoRo}, that in detectors the backreaction from the field to the detector is negligible.  The inclusion of backreaction leads to effective open-system dynamics  with dissipation and noise, which are appropriate to probes of the field, rather than detectors.
For detectors, the leading order terms in perturbation theory provide an accurate characterization of detection probabilities \cite{AnSav11}. For a large sampling of applications of the UdW detectors, see Ref. \cite{HLL12}.

UdW detector models are simple and practical. Their main limitation is that the detector degrees of freedom are not described by  a QFT. As a result, they may lead to non-causal signals in systems that involve more than one detector \cite{superluminal, RelQOp2}.

Finally, we must note the analyses of the measurement process in relation to causality and locality within algebraic QFT \cite{HeKr, OkOz, Dop, FeVe}. In particular, Ref. \cite{FeVe} considers a system and a probe, both described by a QFT. The two field systems are independent and they  interact within a bounded spacetime region. Their interaction can be described by an S matrix, thus leading to correlations between observables on the system and records on the probe. One can define probabilities for the latter using the measurement theory for some  operators that are well defined on the probe Hilbert space. This method is quite general and it avoids the problems of ideal measurements that were described in Sec. 3.3. However, it has not yet been developed into a practical tool, leading to
concrete physical predictions, for example,   photodetection probabilities.

\section{The Quantum Temporal Probabilities program}

\subsection{Key ideas}
In this section, we describe the QTP approach to QFT measurements, which we have been developing for a number of years \cite{QTP1, QTP2, QTP3, QTP4, QTP5}.
The QTP approach aims to construct a framework for measurements that (i) works, in principle, for any QFT, (ii) allows for a first-principles investigation of all issues of causality and locality in relation to QFT measurements, and (iii) it can be directly applied to experiments of current interest.

The main points in the QTP approach are the following.
\begin{enumerate}
\item A measurement requires the interaction between a quantum system and a measurement apparatus. The latter must be a macroscopic system that behaves effectively as a classical. This means that the pointer variables must be highly coarse-grained observables and that the histories of measurement outcomes must satisfy appropriate decoherence criteria, as established in the decoherent-histories approach to quantum mechanics \cite{histories}.

    \item Physical measurements are localized in space and in time. For example, an elementary solid state detector has a specific location in a lab and it records a particle at a specific moment of time that is determined with finite accuracy. In principle, both position and time can be random variables. For example, when directing a single particle towards an array of detectors, both the elementary detector that records the particle (i.e., the location of the detection record) and the time of recording vary from one run of the experiments to the other. Hence, the predictions of the theory must be expressed in terms of probability densities
        \bey
        P(X_1, \lambda_1; X_2, \lambda_2, \ldots, X_n, \lambda_n),
        \eey
for multiple detection events. Here, $X_i$ are spacetime points, $\lambda_i$ stand for any other observable that is being measured and $P$ is a probability density with respect to both $X_i$ and $\lambda_i$.

\item Hence, QFT measurements require the construction of probabilities for  observables that are intrinsically temporal. This is why QTP grew out of a formalism for the description of the time-of-arrival in quantum theory \cite{AnSav06}. The key idea  is to distinguish between the time parameter of Schr\"odinger equation from the time variable associated to  particle detection \cite{Savvidou, Sav09}. The latter time variable is then treated as a  macroscopic quasi-classical one  associated to the detector degrees of freedom. Hence, although  the detector is described  in microscopic scales by quantum theory, its macroscopic records are expressed  in terms of classical spacetime coordinates.

\end{enumerate}

\subsection{The probability formulas}
QTP expresses quantum probabilities in terms of QFT correlation functions. The probability density associated with n measurement events is a linear functional of a 2n unequal-time field correlation function. For example, the probability density
  $P(X)$ that a particle is detected at spacetime point $X$ is
   of the form
\bey
			P(X)=  \int d^4 \xi K(Y)  G_2 (X -\frac{1}{2} \xi, X + \frac{1}{2} \xi),     \label{QTPprob}
\eey
where $K(\xi)$   is a kernel that contains all information about the structure and state of motion of the detector, and
\bey
			G_2 (X,X' )= \langle \psi|\hat{O}(X)\hat{O}(X')|\psi\rangle,
\eey
 is the (two-point) Wightman function associated to a local composite operator $\hat{O}(X)$ on the Hilbert space of the quantum field. The composite operator describes the coupling of the system to the detector.

 Eq. (\ref{QTPprob}) can be derived from a first-principles decoherent histories analysis of the measurement process, in which the spacetime coordinate of the measurement event is a coarse-grained macroscopic observable---see \cite{QTP3} and also \cite{QTPunpubl} for full details in the derivation. Eq. (\ref{QTPprob}) is the leading-order term in a perturbation expansion with respect to the system-apparatus coupling. Like in Glauber's theory, it is useful to treat this term as the signal of our measurement theory, and to treat higher order terms as noise rather than as corrections.
A heuristic derivation of  Eq. (\ref{QTPprob}) is contained in the Appendix.

Eq. (\ref{QTPprob}) defines a Positive Operator Valued Measure (POVM) on the field Hilbert space---a POVM is the most general way of rigorously defining quantum probabilities associated to measurements. From this POVM, we can derive a novel relativistic time-energy uncertainty relation\cite{QTP3} , localization measures for massive relativistic particles, and to define tunneling time for relativistic particles\cite{QTP4}.

 It is important to emphasize that the kernels K are not arbitrary functions, but they are derived from the physical characteristics of the detector. For example, the kernel $K(\xi)$ that appears in Eq. (\ref{QTPprob})is defined as			
\bey
		 K(\xi)= \langle \omega|e^{-i p \cdot \xi}|\omega \rangle,
\eey
where $p$ is the four-momentum operator for the detector degrees of freedom and $|\omega \rangle$ an appropriate vector state on the Hilbert space of the detector.  The fact that $K$ has this specific form is crucial for proving important properties of the probability distribution Eq. (\ref{QTPprob}).

For $n$ detection events at spacetime points $X_1, X_2, \ldots, X_n$, the QTP probability density is
\bey
P_n(X_1, X_2, \ldots, X_n )= \int d^4 Y_1 d^4 Y_2 \ldots d^4Y_n  K_1 (Y_1) K_2 (Y_2 ) \ldots K_n(Y_n) \nonumber \\
G_{2n} (X_1-\frac{1}{2} Y_1, X_2-\frac{1}{2} Y_2, \ldots, X_n - \frac{1}{2}Y_n; X_1+ \frac{1}{2} Y_1, X_2 + \frac{1}{2} Y_2, \ldots, X_n + \frac{1}{2}Y_n), \label{QTP4}
\eey
where $K_1, K_2, \ldots K_n$ are kernels associated to the $n$ detectors, and
\bey
G_{2n} (X_1, X_2, \ldots, X_n;   X'_1, X'_2, \ldots, X'_n) =
\nonumber \\
\langle \psi|A[\hat{O}(X_1) \hat{O}(X_2) \ldots \hat{O}(X_n) ]T[\hat{O}(X_1') \hat{O}(X_2') \ldots \hat{O}(X_n') ]|\psi\rangle, \label{G2n}
\eey
is a $2n$-point function of which the first $n$ indices are anti-time-ordered and the last $n$ indices are time-ordered. Versions of Eq. (\ref{QTP4}) gave been employed  for the analysis of temporal aspects of quantum entanglement in non-relativistic systems \cite{QTP2}, quasi-classical paths in quantum tunneling \cite{QTP5}, and correlations in Hawking radiation \cite{QTP6}.

Correlation functions like Eq. (\ref{G2n}) do not appear in S-matrix theory, as they describe real-time causal evolution. They involve both time-ordered and anti-time-ordered entries, as in the so-called Schwinger-Keldysh formalism \cite{CTP}, now broadly used in many areas of physics from condensed matter physics to cosmology \cite{NEQFT}. The Schwinger-Keldysh formalism has close links to histories theory, for example, the associated generating functional is the functional Fourier transform of the decoherence functional in the associated histories theory \cite{An01}.

The key feature of the QTP probability formula is the explicit relation between a macroscopic notion of causality and the cluster decomposition of the correlation functions (\ref{G2n}). Consider, for simplicity, the case $n = 2$. If $X_1$ and $X_2$ are spacelike separated, then we expect that
\bey
P_2(X_1, X_2) = P_1(X_1) P_2(X_2) \label{seploc}
\eey
Let us assume that the kernels $K_1$ and $K_2$ vanish outside a region of compact support ${\cal C}$. For any $Y_1, Y_2 \in {\cal C}$, we have two clusters: one consisting of points  $X_1 \pm\frac{1}{2}Y_1$ and the other consisting of points $X_2 \pm\frac{1}{2}Y_2$. If the two clusters are spacelike separated for all $Y_1, Y_2 \in {\cal C}$, then the cluster decomposition property for $G_{2n}$ implies the locality condition (\ref{seploc}). Hence, if $X_1$ and $X_2$ have a strong spacelike separation, i.e., $X_1 - X_2$ is sufficiently far from the lightcone, causality is expected.

However, the problem is that physical kernels $K_1$ and $K_2$ may not be of finite support, possibly  leading to small violations of Eq. (\ref{seploc}). This does not necessarily imply a violation of causality, because Eq. (\ref{seploc}) does not take into account the noise from higher order interaction processes. A violation of causality would require a faster-than-light signal. We conjecture that, with appropriate constraints on the kernels $K_1$ and $K_2$, most, if not all, apparent violations of causality will be so small as to be hidden by the noise from higher-order processes.

\section{Conclusions}
We have argued that the extension of quantum information theory to relativistic systems---including quantum gravity---requires the formulation of a consistent and practicable quantum measurement theory for QFT. We presented the challenges that must be overcome by such a theory, and we presented the main ideas of  the QTP program that aims to provide such a theory.

This issue of  particular importance for deep space experiments, which provide
a new frontier for Quantum Information Science and
for fundamental physics, especially quantum foundations. These experiments will allow us to measure quantum correlations at distances
of the order of $10^5$km and for detectors with large relative velocities. The Deep Space
Quantum Link (DSQL) mission  envisions such experiments with photons that involve
either Earth-satellite- or intra-satellite communications \cite{DSQL}. These experiments will allow us to test for the first time the foundations of QFT in relation to causality and locality, and by extension to test between different photodetection models appropriate for this novel regime.

 Deep space experiments will also enable us to understand
 the influence of relativistic effects on quantum resources, like entanglement. These effects include relative motion of detectors, retarded propagation
at long distance, distinction between timelike and spacelike correlations and gravity
gradients. It will also allow us to consider novel types of quantum correlations that
are more ”relativistic” in nature, e.g., correlations between temporal variables and
qubit variables. A consistent QFT measurement theory ought to provide precise predictions for all effects of this type.

\section*{Acknowledgements}
Research was supported by   grant  JSF-19-07-0001 from the Julian Schwinger Foundation. CA would like to thank Bei-Lok Hu, Maria Papageorgiou and Juan de Ramon for discussions on the topic.

\begin{appendix}
\section{An elementary derivation of the QTP probability formula}
Consider a QFT described by a Hilbert space ${\cal F}$, which carries a unitary representation of the Poincar\'e group.  Working in the Heisenberg picture, we denote the fields by $\hat{\phi}_r(X)$.

The apparatus is a physical system described by a Hilbert space ${\cal K}$. The apparatus follows a world tube ${\cal W}$ in Minkowski spacetime. We assume that the size of the apparatus is finite, but still much larger than the scale of microscopic dynamics (the atomic scale usually), so that it is meaningful to treat the ground state $|\Omega\rangle$ of the apparatus as invariant under spatial translations generated by the energy-momentum vector $\hat{p}^{\mu}$ for the apparatus. We also assume that there is a gap between the ground state and any excited state of the apparatus.

We assume a coupling between the system and the apparatus with support only in a small spacetime region around a point $X$. The finite spacetime extent of the coupling is arguably unphysical, but it serves to mimic the effect of a detection record localized at $X$. To leading order in perturbation theory, the resulting probabilities coincide with those obtained from a more detailed modeling of the measurement act through localized POVMs that are correlated to position \cite{QTP3}. The interaction term is
\bey
\hat{V}_X = \int f(X, Y) \hat{O}_a(Y) \otimes \hat{J}^a(Y),
\eey
where $\hat{O}_a(X)$ is a composite operator on ${\cal F}$ that is local with respect to the field $\hat{\phi}_r(X)$ and $a$ is a label that includes both spacetime and internal indices. The current operators $\hat{J}^a(X)$ are defined on ${\cal K}$. We must assume that $\langle \omega|\hat{J}^a(X)|\omega\rangle = 0$

 The switching functions $f(X, Y)$ are dimensionless, they depend on the motion of the apparatus and they vanish outside the apparatus world tube and at times that the interaction is switched-off. For an apparatus that is static in a specific inertial frame, we can simply choose $f(X, Y) = f(X - Y)$.

For an elementary detector, i.e., a detector that supports a single detection record, It is convenient to work with a Gaussian $f$; for $X = (t, {\bf x})$,
\bey
f(t, {\bf x}) = \exp \left( - \frac{t^2}{2\delta_t^2} - \frac{|{\bf x}|^2}{2 \delta_x^2}\right),
\eey
where $\delta_t$ is the duration of the interaction and $\delta_x$ is the size of the detector. Both are assumed to be macroscopic scales, in order to make contact with the detailed measurement theory employed in QTP, in particular with the existence of macroscopic records of observation \cite{QTP3}.

The Gaussian switching functions satisfy a useful identity
\bey
f(X) f(X') = f^2\left(\frac{X+X'}{2}\right) \sqrt{f}(X - X'). \label{gauidty}
\eey
The volume $\upsilon$ of the spacetime region in which the field-apparatus interaction is switched on is
\bey
\upsilon = \int dt d^3 x f(t, {\bf x})^2 = \pi^2 \delta_t \delta_x^3.
\eey
We note that the function $F(X): = \frac{1}{\upsilon} d^2(X)$ is a normalized probability density on $M$.

The probability $\mbox{Prob}(X)$  that the detector becomes excited after the interaction is completed is, to leading order in perturbation theory,
\bey
\mbox{Prob}(X) = \int d^4Y_1 d^4Y_2  f(X - Y_1) f(X - Y_2) G_{ab}(Y_1, Y_2) \langle \Omega|\hat{J}^a(Y_1) \left(\hat{I} - |0\rangle \langle 0|\right) \hat{J}^b(Y_2)|\Omega\rangle, \label{probX}
\eey
where
\bey
G_{ab}(X, X')  = \langle \psi|\hat{O}_a(X) \hat{O}_b(X')|\psi\rangle,
\eey
is a correlation function for the quantum field. We have assumed a factorized initial state $|\psi\rangle \otimes |\Omega\rangle$ for the total system; $|\psi\rangle$ is an arbitrary state for the field.

Let $X = 0$ be a reference point on the world-tube of the apparatus. Then, we can write $\hat{J}^{a}(X) = e^{-i \hat{p} \cdot X} \hat{J}^a(0) e^{i \hat{p} \cdot X}$. For a translation-invariant $|0\rangle$, $\langle \Omega|\hat{J}^a(Y_1) \left(\hat{I} - |\Omega\rangle \langle \Omega|\right) \hat{J}^b(Y_2)|\Omega\rangle = K^{ab}(Y_2 - Y_1)$, where
\bey
K^{ab}(Y) := \langle \Omega| \hat{J}^a(0) e^{-i\hat{p}\cdot Y}\hat{J}^b(0)|\Omega\rangle - \langle \Omega |\hat{J}^a(0)|\Omega\rangle \langle \Omega |\hat{J}^b(0)|\Omega\rangle.
\eey
The probability $\mbox{Prob}(X)$ of Eq. (\ref{probX}) is not a density with respect to $X$; $X$ appears as a parameter of the switching function.    In classical probability theory, we could define an unnormalized probability density $W(X)$ with respect to $X$ by dividing $\mbox{Prob}(X)$ with the effective spacetime volume $\upsilon$. Then, using Eq. (\ref{gauidty}) we find,
\begin{eqnarray}
W(X) = \int d^4X F(X - X') P(X'), \label{Prob0b}
\end{eqnarray}
where
\bey
P(X) = \int d^4 \xi  \sqrt{f}(\xi) K^{ab}(Y) G_{ab}(X - \frac{1}{2}\xi, X +\frac{1}{2}\xi).
\eey
The  definition (\ref{Prob0b}) of a spacetime density with respect to time is not rigorous for quantum probabilities, because it involves the combination of probabilities defined with respect to different experimental set-ups, i.e., different switching functions for the Hamiltonians. There are numerous theorems asserting that such combinations may not be acceptable in quantum theorem.
Nonetheless,   Eq. (\ref{Prob0b}) can  be derived as a genuine probability density in the context of the QTP method \cite{QTP1, QTP2, QTP3}, as long as we restrict to the leading order of perturbation theory.

QTP leads to different predictions from the method presented here at higher orders of perturbation theory.   In QTP, the interaction is present at all times, as it should be in any first-principles derivation. The
smearing functions $f(X)$ are not interpreted in terms of a switching-on of the interaction,
 but they describe the {\em sampling} of a temporal observable associated to a point $X$ of the apparatus' world tube through a detection record. Sampling functions in QTP they incorporate the {\em  coarse-graining} necessary for the definition of classicalized pointer variables.

The probability distribution $W(X) $ is the convolution of $P(X)$ with the probability density $F(X)$ that incorporates the accuracy of our measurements classically. If $P(X)$ is non-negative and the scale of variation in $X$ is much larger than both $\delta_t$ and $\delta_x$, we can treat $P(X)$  as a finer-grained version of $W(X)$ and employ this as our probability density for detection.

The kernel $K^{ab}(\xi)$ is typically characterized by some correlation length-scale $\ell$ and some correlation time-scale $\tau$, such that $ K^{ab}(\xi)  \simeq 0$ if $|t(\xi)| >> \tau $ or  $|{\bf x}(\xi)|>> \ell$.
Both scales $\ell$ and $\tau$ are microscopic and characterize the constituents of the apparatus and their dynamics. If $\ell << \sigma_x$ and $\tau << \sigma_t$, then $K^{ab}(\xi) \sqrt{f}(\xi) \simeq K^{ab}(\xi)$ and the QTP probability formula simplifies to
\bey
P(X) = \int d^4 \xi    K^{ab}(Y) G_{ab}(X - \frac{1}{2}\xi, X +\frac{1}{2}\xi).
\eey

\end{appendix}

\end{document}